\documentclass{aa}
\usepackage{graphics}
\def\a{{$\alpha$}}

\def\gsnr{{G~65.3+5.7}}
\newcommand{\h}{$^{\rm h}$}
\newcommand{\m}{$^{\rm m}$}
\newcommand{\s}{$^{\rm s}$}
\newcommand{\dd}{$\delta$}
\newcommand{\ha}{\rm H$\alpha$}
\newcommand{\hbeta}{\rm H$\beta$}
\newcommand{\HII}{\ion{H}{ii}}
\newcommand{\hnii}{{\rm H}$\alpha+[$\ion{N}{ii}$]$}
\newcommand{\nii}{$[$\ion{N}{ii}$]$}
\newcommand{\sii}{$[$\ion{S}{ii}$]$}
\newcommand{\oii}{$[$\ion{O}{ii}$]$}
\newcommand{\oi}{$[$\ion{O}{i}$]$}
\newcommand{\oiii}{$[$\ion{O}{iii}$]$}

\newcommand{\et}{et al.}
\newcommand{\flux}{$10^{-17}$ erg s$^{-1}$ cm$^{-2}$ arcsec$^{-2}$}
\newcommand{\dens}{\rm cm$^{-3}$}
\newcommand{\sdens}{\rm cm$^{-2}$}
\newcommand{\vel}{\rm km s$^{-1}$}
\begin{document}

%
\title{Deep optical observations of \gsnr}

\author{F. Mavromatakis\inst{1}
\and P. Boumis\inst{1}
\and J. Papamastorakis \inst{1,2}
\and J. Ventura \inst{1}}
\offprints{F. Mavromatakis,\email{fotis@physics.uoc.gr}}
\authorrunning{F. Mavromatakis et al.}
\titlerunning{Optical observations of \gsnr}
\institute{
University of Crete, Physics Department, P.O. Box 2208, 710 03 Heraklion, 
Crete, Greece 
\and Foundation for Research and Technology-Hellas, P.O. Box 1527, 
711 10 Heraklion, Crete, Greece}
\date{Received 12 February 2002 / Accepted 4 April 2002}

\abstract{We present the first CCD mosaic of the supernova remnant 
\gsnr\ in the optical emission lines of \oii\ and \oiii. 
The new images reveal several diffuse and filamentary structures 
both inside and outside the extent of the remnant as defined by its 
X--ray and radio emission. 
The medium ionization line of \oiii 5007 \AA\ provides the sharpest view to 
the system, while the remnant appears less filamentary in the emission 
line of \oii. There are significant morphological differences between the 
two images strongly suggesting the presence of incomplete shock structures. 
Deep long--slit spectra were taken at several different positions of \gsnr. 
All spectra originate from shock heated gas, while the majority of them 
is characterized by large \oiii/\hbeta\ ratios. The sulfur line ratios indicate 
electron densities below $\sim$200 \dens, while estimates of the shock 
velocities lie in the range of 90--140 \vel. Finally, the observed variations 
of the \ha/\hbeta\ ratios may reflect the presence of intrinsic absorption 
affecting the optical spectra.
}
\maketitle
\section{Introduction}
Gull \et\ (\cite{gul77}) reported the detection of a new supernova 
remnant in Cygnus during an optical survey with narrow band filters. In
particular, they obtained an \oiii\ photograph showing the existence  
of a $\sim$3\degr.3$\times$4\degr.0 filamentary shell. 
Due to its large extent, high resolution
imagery has been performed only in selected regions of the remnant 
(Fesen \et\ \cite{fes83} and references therein). 
Subsequent, optical spectra obtained by Sabbadin \& D'Odorico (\cite{sab76}) 
and Fesen \et\  (\cite{fes85}) suggested emission from shock heated gas. 
A stratification among lower and medium ionization lines has been reported 
by Sitnik \et\ (\cite{sit83}) for another area of \gsnr. 
Fesen \et\ (\cite{fes85}) obtained long--slit spectra at two different
positions of \gsnr\ showing strong sulfur line emission relative to \ha\ 
and especially one of the spectra was characterized by very strong \oiii\ 
emission compared to \hbeta\ ($\sim$40). High \oiii/\hbeta\ ratios have  
been observed in other remnants as well and imply the presence of incomplete 
recombination zones (e.g. Raymond \et\ \cite{ray88}). 
Radio observations at 1420 MHz (Reich \et\ \cite{rei79}) confirmed the 
non-thermal nature of the radio emission from this shell and found a 
spectral index of $\sim$0.61.
Current estimates of the linear diameter and the distance to the remnant 
are $\sim$70 pc and 0.9--1.0 kpc, respectively (Gull \et\ \cite{gul77}; 
Reich \et\ \cite{rei79}; Rosado \cite{ros81}; Sitnik \et\  \cite{sit83}). 
Soft X--ray observations show that \gsnr\ is a weak source of X--ray emission  
and imply a shock velocity in the range of a few hundred \vel\ and an age 
of $\sim$ 20000 yrs (Snyder \et\ \cite{sny78}; Mason \et\ \cite{mas79}; 
Seward \cite{sew90}). \gsnr\ has also been observed by ROSAT both during 
the All-Sky survey (Aschenbach \cite{asc94}) and in pointed mode (Lu \& 
Aschenbach \cite{lu02}).  The analysis of these data showed that the 
X--ray emission is clumpy and that the forward shock is traveling with 
a velocity of $\sim$400 \vel\ about 26000 yrs after the supernova 
explosion, while the explosion energy and the interstellar medium density 
are $\sim$2$\times$10$^{50}$ erg and $\sim$0.02 \dens, respectively. 
These parameters were derived assuming a distance to the remnant of 1 kpc.  
Gorham \et\  (\cite{gor96}) have performed a pulsar search of several known 
SNRs at 430 and 1420 MHz. Their observations were sensitive to pulsars with 
periods greater than 1 ms, and flux densities as low as 0.2 mJy but no pulsar was 
found to be associated with \gsnr\ down to these limits.
\par
In order to study in detail this extended and nearby supernova remnant 
we performed deep optical imaging observations in the low and medium 
ionization lines of \oii 3727 \AA\ and \oiii 5007 \AA. 
We have also obtained flux calibrated images of a specific field of 
\gsnr\ in \hnii, \sii, \oii\ and \oiii. The imaging observations are 
supplemented by deep optical long--slit spectra at 8 different locations
of \gsnr. Information about the observations and the data 
reduction is given in Sect. 2. In Sect. 3, 4 and 5 we present the results of 
our imaging observations and the long--slit spectra. 
Finally, in Sect. 6 we discuss the physical properties of the remnant.

\section{Observations}
\subsection{Optical images}
The observations presented in this paper were performed with the 0.3 m 
Schmidt Cassegrain telescope at Skinakas Observatory, Crete, Greece. 
The supernova remnant \object{G~65.3+5.7} was observed on July 22, 23 and 24, 2001. 
A 1024 $\times$ 1024 SITe CCD was used  which 
had a pixel size of 24 $\mu$m resulting in a 89\arcmin\ $\times$ 89\arcmin\ 
field of view and an image scale of 5\arcsec\ per pixel. 
Several pointings were performed in order to cover the extended 
area of \gsnr. Each of the nine (9) different fields was 
observed for 2400 s. All fields were projected to a common origin on 
the sky and were subsequently combined to create the final mosaics 
in \oii\ and \oiii. The astrometric calculations utilized the HST 
Guide star catalog (Lasker \et\ \cite{las99}).
In addition, we obtained flux calibrated images of a specific field 
of \gsnr\ in \hnii, \sii, \oii\ and \oiii. The exposure time of these images
is 1800 s with the exception of the \oii\ image where we obtained two images,
1800 s each. This observation was performed in July 13, 1999 with the 
same telescope and hardware setup as the latest observations.  
The filter characteristics can be found in Mavromatakis \et\ (\cite{mav00}).
All coordinates quoted in this work refer to epoch 2000.
  \begin {figure*}
   \resizebox{\hsize}{!}{\includegraphics{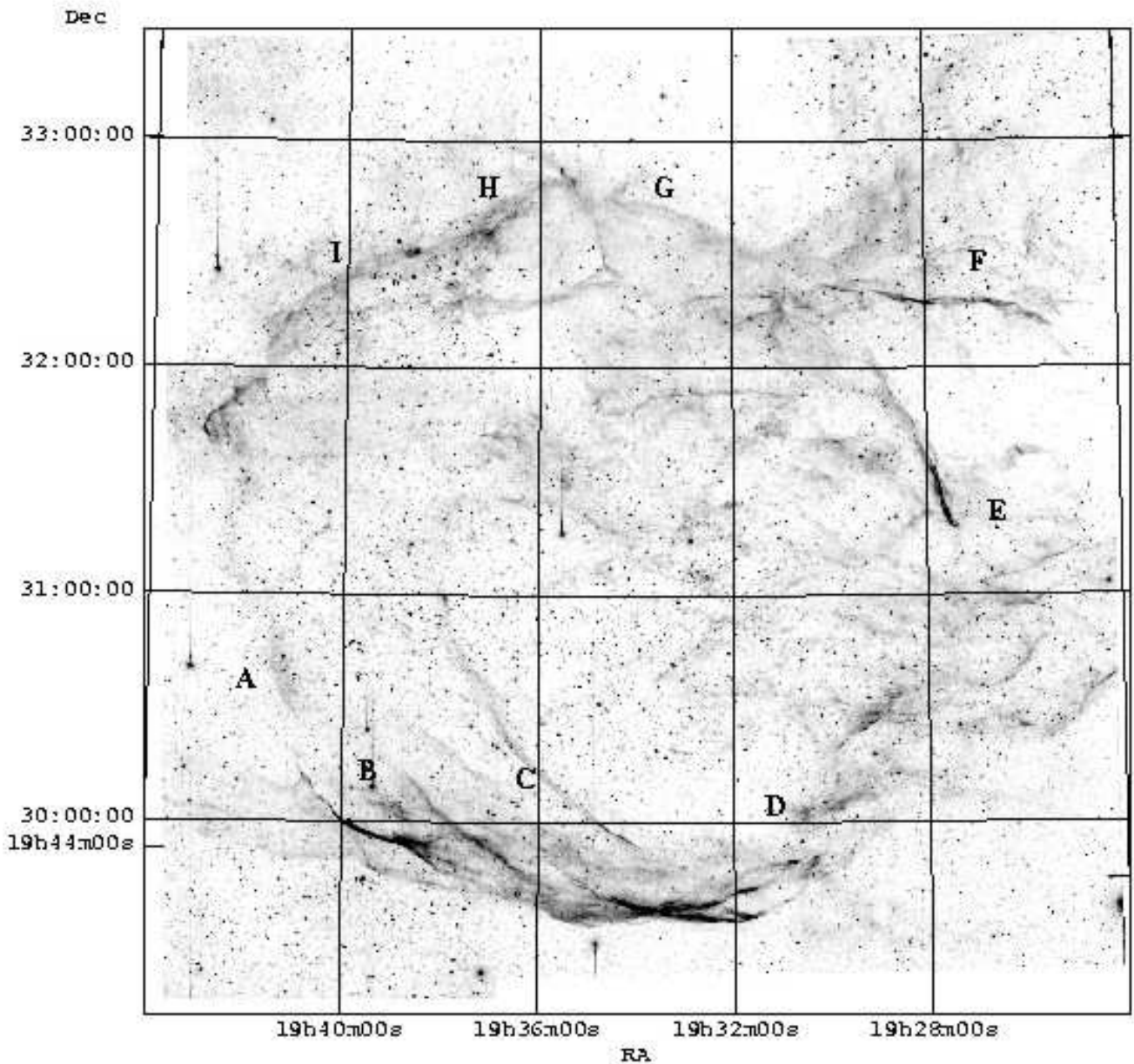}}
    \caption{ The 4\degr.3 square field covering the supernova remnant 
    \gsnr\ imaged in the low ionization line of \oii 3727 \AA. The
     letters define the areas which are discussed in the text in more
     detail. 
    } 
     \label{fig01}
  \end{figure*}
\par
We employed standard IRAF and MIDAS routines for the reduction of the data. 
Individual frames were bias subtracted and flat-field corrected using 
well exposed twilight flat-fields. The spectrophotometric standard stars 
HR5501, HR8634 and HR7950 were used for absolute flux calibration in the 1999
observation of \gsnr.

  \begin{table}
      \caption[]{Log of the spectroscopic observations}
         \label{slits}
\begin{flushleft}
\begin{tabular}{lllllll}
            \noalign{\smallskip}
\hline
 Slit centers &                               &No of spectra  \cr
\hline
     $\alpha$  (epoch 2000) & $\delta$ (epoch 2000)          & (Exp. time)   \cr
\hline 
19\h42\m55.3\s & 31\degr43\arcmin39\arcsec (Area 1) & 2$^{\rm a}$ (3900)$^{\rm b}$ \cr
\hline
19\h40\m00.1\s & 29\degr59\arcmin59\arcsec (Area 2) & 2$^{\rm a}$ (3900)$^{\rm b}$ \cr
\hline
19\h31\m45.0\s & 29\degr34\arcmin55\arcsec (Area 3) & 2$^{\rm a}$ (3900)$^{\rm b}$ \cr
\hline
19\h27\m45.7\s & 31\degr30\arcmin10\arcsec (Area 4) & 1$^{\rm a}$ (3600)$^{\rm b}$ \cr
\hline
19\h28\m00.9\s & 32\degr16\arcmin59\arcsec (Area 5) & 2$^{\rm a}$ (3900)$^{\rm b}$ \cr
\hline
19\h35\m34.0\s & 32\degr54\arcmin32\arcsec (Area 6) & 2$^{\rm a}$ (3900)$^{\rm b}$ \cr
\hline
19\h38\m43.0\s & 32\degr28\arcmin48\arcsec (Area 7) & 2$^{\rm a}$ (3900)$^{\rm b}$ \cr
\hline
19\h35\m59.2\s & 30\degr12\arcmin35\arcsec (Area 8)& 3$^{\rm a}$ (2000)$^{\rm b}$ \cr
\hline
\end{tabular}
\end{flushleft}
${\rm ^a}$ Number of spectra obtained \\\
${\rm ^b}$ Exposure time of individual spectra in sec\\\
   \end{table}
 \begin{table*}
        \caption[]{Relative line fluxes}
         \label{sfluxes}
         \begin{flushleft}
         \begin{tabular}{lllllllll}
     \hline
\noalign{\smallskip}
           & area 1a      & area 1b   & area 1c& area 2   & area 3a  &  area 3b & area 4 & area 5 \cr  
\hline  
Line (\AA)  &             &           &         &           &         &           &        &   \cr  
\hline  
4861 \hbeta\ & 22 (8)    & 20 (14)    & 13  (5)  & 20 (6)   & 27 (23)  & 25 (26)  & 22 (11)& 28 (32)\cr  
 \hline  
4959 \oiii\  & 240 (95)  & 417 (104)  & 360 (103)& 128 (38) & 84 (93)  & 38 (33) & 52 (25)& 50 (74) \cr  
\hline  
5007 \oiii\  & 757 (197) & 1273 (272)& 1074 (153)& 378 (114)& 263 (166)&123 (87) & 170(70)& 158 (173)\cr  
\hline  
6300 \oi\   &	--	  & --         &  --     &  --       & 6 (18)  & 16 (35)  & 12 (11)& 8 (11)\cr
\hline
6364 \oi\   &   --        &  --        &  --     &  --       &  --     & 6 (8)    & 4 (4)  & 2 (3) \cr
\hline
6548 \nii\  & 30 (15)    & 35 (27)   & 37 (15)   &  40 (24)& 26 (54)   & 25 (44)  & 37 (32)& 30 (66) \cr  
\hline  
6563 \ha\   & 100 (39)   & 100 (64)  & 100 (45)  &  100 (60)& 100 (110)& 100 (112)& 100 (77)& 100 (156)\cr  
\hline  
6584 \nii\  & 111 (54)   & 116 (71)  & 130 (47)  &  125 (75)& 87 (109) & 82 (102) & 114 (78)& 93 (152) \cr  
\hline  
6716 \sii $_1$  & 74 (32)    & 87 (56)   & 103 (34)  &  75 (50) & 74 (93)  & 56 (121) & 96 (74) & 61 (121) \cr  
\hline  
6731 \sii $_2$  & 55 (24)    & 66 (42)   & 76 (25)   &  56 (37) & 52 (66)  & 41 (88)  & 67 (52) & 43 (83) \cr  
\hline
\hline  
Absolute \ha\ & 6.7      & 9.3       & 10.7      & 20.6     & 16.3      &  22.1     & 10.4    &17.3  \cr  
\hline  
\ha/\hbeta\    & 4.5 (8)  & 5.0 (14)   & 7.7 (5)  & 5.0 (6)  & 3.7 (23) & 4.0 (26)  &4.5 (11)  &3.6 (32) \cr  
\hline
  c         & 0.5 (4)  & 0.7 (7)    & 1.2 (5)   & 0.7 (3)  & 0.28 (5)  & 0.38 (7) & 0.55 (6) & 0.24 (6) \cr  	  
\hline
\oiii /\hbeta\ & 45 (8)   & 85 (14)    & 110 (5)   & 25 (6)   & 13 (23)   & 6.4 (25)  &10 (11) &7.4 (32)\cr  
\hline  
\sii/\ha\     & 1.29 (28)  & 1.53 (47) & 1.79 (31) & 1.31 (43)& 1.26 (90) & 0.97 (95) &1.63 (58) &1.04 (106) \cr  
\hline   
\sii $_1$/\sii $_2$    & 1.34 (19)  & 1.32 (34) & 1.35 (20) & 1.34 (30)& 1.42 (54) & 1.37 (71) &1.43 (43) &1.42 (68)\cr  
\hline
n$_{\rm e}$ (\dens)& $<$ 150& $<$ 140   & $<$ 140   & $<$ 130  & $<$ 45     & $<$ 70    & $<$ 35    & $<$ 30 \cr
\hline
\end{tabular}
\begin{tabular}{lllll}
   & & & & \cr  

           & area 6     & area 7   & area 8a & area 8b\cr  
\hline  
Line (\AA)  &            &          &         &         \cr  
\hline  
4861 \hbeta\ & --         & 24 (18)  & --      & --      \cr  
\hline  
4959 \oiii\  & 135 (86)  & 104 (69)  & 378 (31)& 690 (48)\cr  
\hline  
5007 \oiii\  & 415 (239) & 326 (150) & 1265 (91)& 2040 (141)\cr  
\hline  
6300 \oi\   &  11 (13)  & 7 (3)     & --       &   --     \cr
\hline
6364 \oi\   &  4 (5)    &  --       & --       &  --      \cr
\hline
6548 \nii\  & 33 (39)   & 27 (16)   &    --    &   --     \cr  
\hline  
6563 \ha\   & 100 (115) & 100 (61)  & 100 (9)  & 100 (10) \cr  
\hline  
6584 \nii\  & 112 (128) & 91 (56)   & 135 (13) & 131 (14) \cr  
\hline  
6716 \sii $_1$  & 95 (115)  & 38 (23)   & 93 (9)   & 84 (8) \cr  
\hline  
6731 \sii $_2$  & 66 (81)   & 30 (19)   & 89 (8)   & 61 (5) \cr  
\hline
\hline  
Absolute \ha & 7.7      & 9.0      & 0.6     & 0.5  \cr  
\hline  
\ha/\hbeta\   &  --       & 4.2 (17)  & --     & --   \cr  
\hline
   c       &   --      & 0.44 (5)   & --    & -- \cr             
\hline  
\oiii /\hbeta\ &  --      & 18 (18) & --    & --   \cr  
\hline  
\sii/\ha\     & 1.61 (88)  & 0.68 (27) & 1.82 (7) & 1.45 (7)\cr  
\hline   
\sii $_1$/\sii $_2$  & 1.44 (66)  & 1.27 (15) & 1.04 (6) & 1.38 (4) \cr  
\hline
n$_{\rm e}$ (\dens) & $<$ 30 & $<$ 170 &           &    \cr
\hline  
\end{tabular}  
\end{flushleft} 
${\rm }$ The numbers next to the areas number indicate different \\
${\rm }$ apertures extracted along the slit\\
${\rm }$ Listed fluxes are a signal to noise weighted average of the 
individual fluxes\\
${\rm }$ and are not corrected for interstellar extinction \\
${\rm }$ Numbers in parentheses represent the signal to noise ratio 
of the quoted fluxes\\
$^{\rm }$ Absolute \ha\ flux in units of \flux\\\
${\rm }$ All relative fluxes normalized to F(\ha)=100
\end{table*}
%
  \begin {figure*}
  \resizebox{\hsize}{!}{\includegraphics{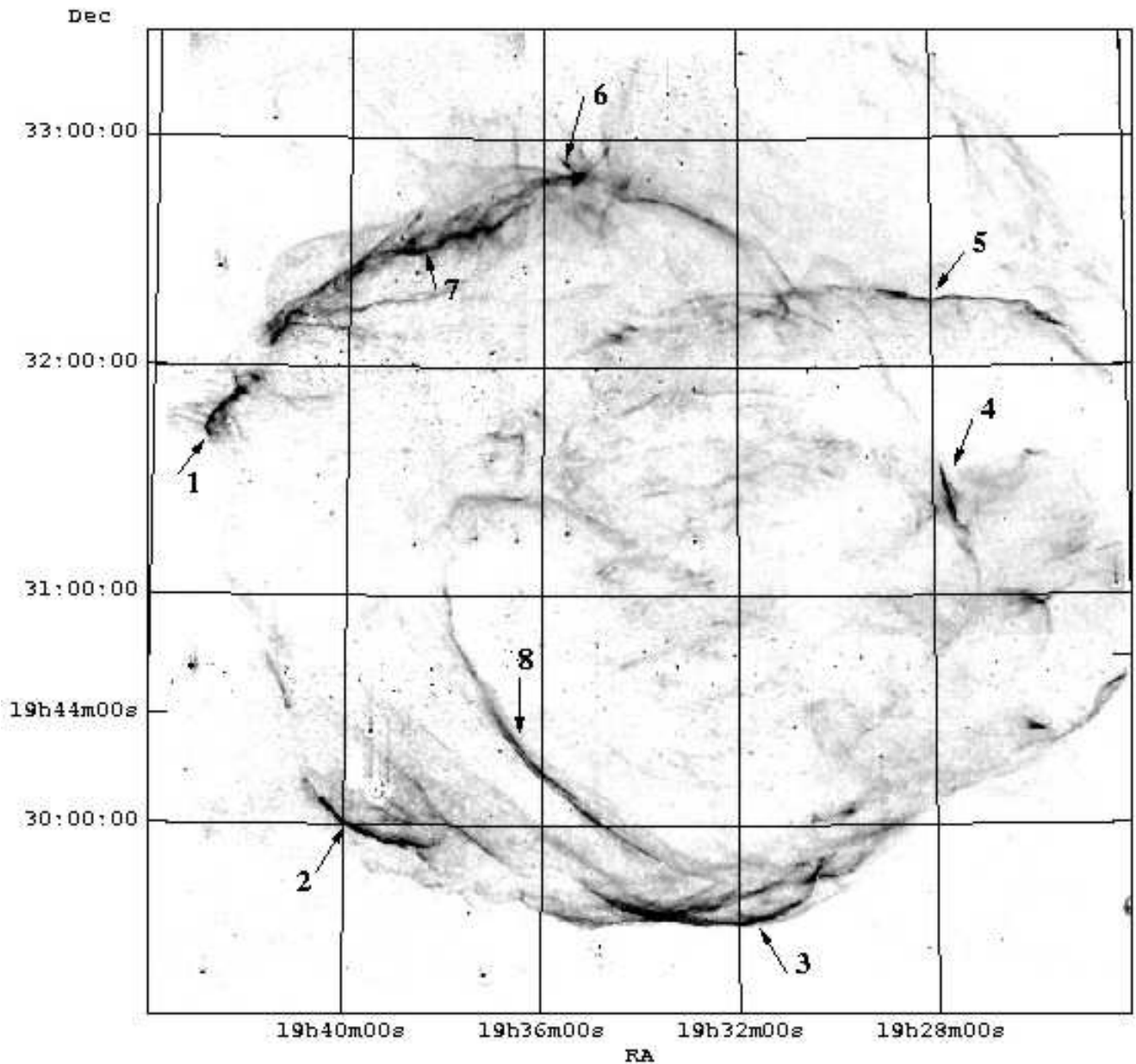}}
  \caption{The remnant \gsnr\ as it appears in the medium ionization 
    line of \oiii 5007 \AA. The filamentary nature is clearly seen in this 
    image with respect to the lower ionization image of \oii\ 
    (Fig. \ref{fig01}). The arrows roughly point to the projected center 
    on the sky of the slit in each location given in Table \ref{slits}, 
    while the numbers designate the individual locations.} 
    \label{fig02}
  \end{figure*}
\subsection{Optical spectra}
Long--slit spectra were obtained on July 15, 1999, and July 24--28, 2001 
using the 1.3 m Ritchey--Cretien telescope at Skinakas Observatory. 
The data were taken with a 1300 line mm$^{-1}$ grating 
and a 800 $\times$ 2000 Site CCD covering the range of 4750 \AA\ -- 6815 \AA.
The slit had a width of 7\farcs7 and a length of 7\arcmin.9 and, in all cases, 
was oriented in the south-north direction. The coordinates of the slit centers 
along with the number of spectra and their individual exposure 
times are given in Table \ref{slits}. 
The spectrophotometric standard stars HR5501, HR7596, HR7950, HR9087, and 
HR718 were observed in order to calibrate the spectra.  
\section{Imaging of \gsnr}
\subsection{The \oii\ and \oiii\ line mosaics}
The wide field covered by \gsnr\ contains a number of extended sources 
like bright nebulae (e.g. \object{LBN 064.90+06.74}; Lynds \cite{lyn65}), 
dark nebulae 
(e.g. \object{LDN 817}, \object{LDN 816}; Lynds \cite{lyn65}) as well as a 
few planetary nebulae (e.g. \object{PN G064.7+05.0}, \object{PN G065.4+03.1}). 
In Fig. \ref{fig01} we show the mosaic in the low ionization line 
of \oii\ along with letters designating the areas which are described in the 
text in detail. 
The \oii\ image appears generally less filamentary than the \oiii\ 
image (Fig. \ref{fig02}). In addition, the former image contains more 
diffuse emission than the latter. 
We start the detailed morphological study of \gsnr\ in these filters from
area A (Fig. \ref{fig01}). In this area diffuse emission is seen in the 
\oii\ image, while strong filamentary emission is detected in the medium 
ionization line image of \oiii. 
The filament is $\sim$16\arcmin\ long and its typical projected 
width is $\sim$1\arcmin. Immediately to the south--west of area A we 
find area B which contains the bright binary \object{$\phi$ Cyg} and 
has been studied by Sitnik \et\ (\cite{sit83}).  They observed a 
narrow field around $\phi$ Cyg ($\sim$6\arcmin.5 $\times$ 6\arcmin.5) and found a 
stratification in the emission from the \oiii, \nii\ and \sii\ lines with the 
\oiii\ emission lying closer to the outer edge of the remnant. 
Here, the \oii\ emission appears more filamentary than the \oiii\ emission.
The $\sim$37\arcmin\ long filament seen in \oii\ is broader in the south and 
becomes narrower as it extends further to the north, while the \oiii\ image 
shows roughly a constant projected thickness.  The projected thickness 
close to the northern tip of the \oii\ filament is $\sim$32\arcsec, while at the
same location the thickness of the \oiii\ filament is $\sim$65\arcsec. 
We also observe here an increasing offset in the outer boundaries of the 
\oii\ and \oiii\ images with the \oiii\ emission leading and the \oii\ emission 
following. 
Further to the west we come across area C for which we have performed 
flux calibrated imaging observations. This area will be discussed in the next 
section separately.     
\par
Significant differences between the low and medium ionization images are 
found in area D. There are two locations where the \oii\ emission is quite 
intense and filamentary, while at the same location the \oiii\ emission is 
diffuse and weak. In this area there is also a location where the inverse 
is observed, i.e. filamentary emission in \oiii\ and diffuse, very weak 
emission in \oii. Area E contains a complex network of filaments and 
diffuse emission in \oii, while the \oiii\ image provides a sharper view 
of the area. LBN 148 (Lynds \cite{lyn65}) is also found in this area, and 
shows up as a 25\arcmin\ long filament in \oii, while it appears much shorter 
in the \oiii\  emission line image ($\sim$18\arcmin). 
Emission in both images is detected further
to the west of LBN~148 and seems to extend outside our field. We note here 
that our coverage to the west marginally reaches \a$\simeq$19\h24\m, while the 
ROSAT data show that emission extends up to \a$\simeq$19\h23\m\ (Lu \&
Aschenbach \cite{lu02}).     
%
  \begin {figure*}
   \resizebox{\hsize}{!}{\includegraphics{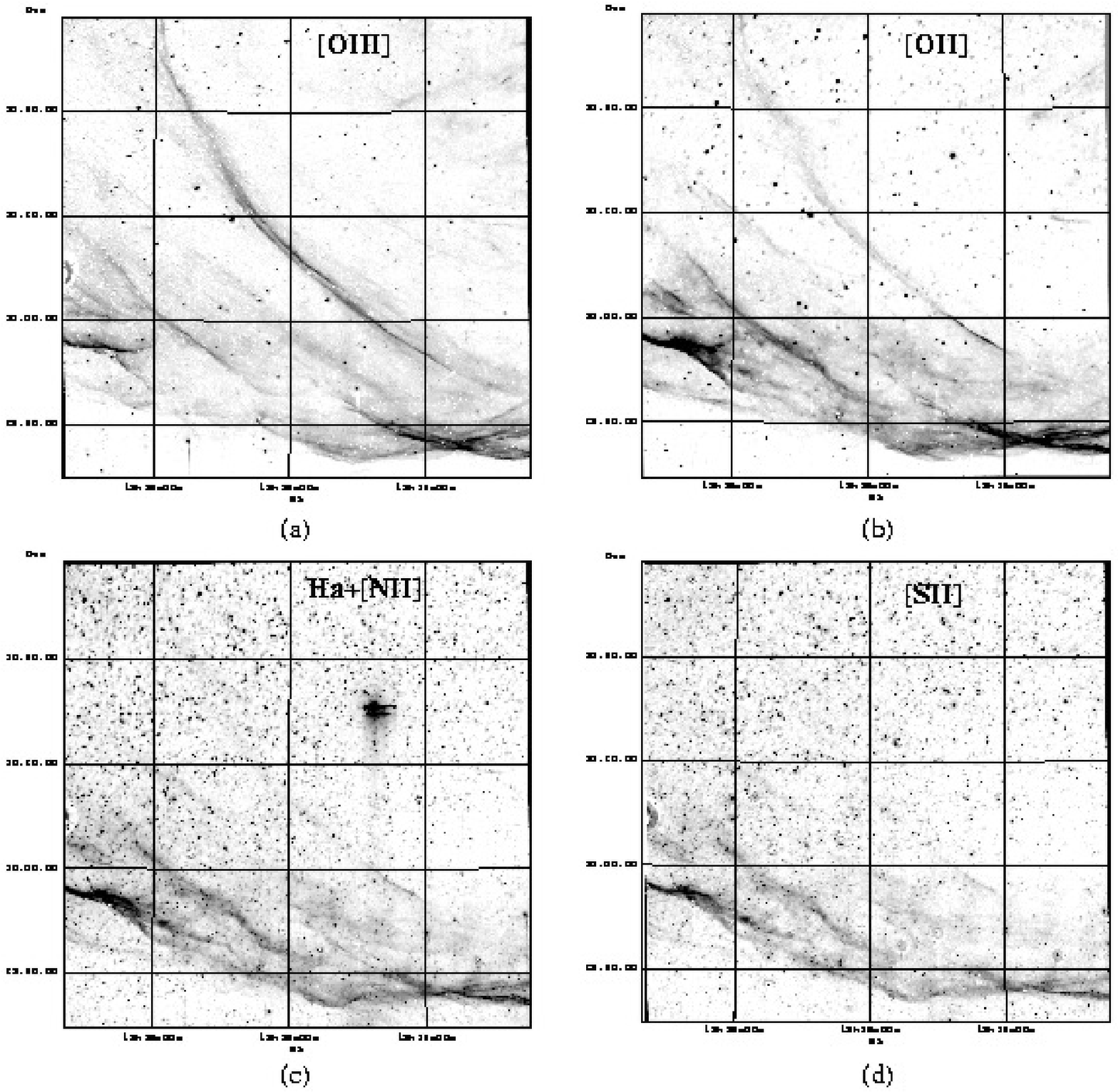}}
   \caption{A 1\degr.5 wide area in the south of \gsnr\ imaged with the 
     \oiii 5007\AA\ (a), \oii 3727 \AA\ (b), \hnii\ (c) and \sii\ (d) 
     filters. The images have been smoothed to suppress the residuals from 
     the imperfect continuum subtraction. The shadings run linearly 
     from 0 to 75 in Fig. \ref{fig03}a, 0 to 40 in Fig. \ref{fig03}b, 
     0 to 80 in Fig. \ref{fig03}c and 0 to 30 in 
     Fig. \ref{fig03}d in units of \flux.} 
     \label{fig03}
   \end{figure*}
\par
In area F we find a long filamentary structure both in \oii\ and \oiii\ 
(Figs. \ref{fig01} and \ref{fig02}) extending from the west edge of our 
field up to the central areas of the remnant. There is a wealth of short 
scale structures to the north of this long filament which are only seen 
in the low ionization image of \oii. 
However, further to the north in the 
\oiii\ image faint, filamentary and diffuse structures are discovered, 
around \a $\simeq$19\h27\m\ and \dd $\simeq$33\degr00\arcmin, which 
seem to extend outside the north boundary of our field and are projected
outside the main body of the soft X--ray emission (Lu \& Aschenbach \cite{lu02}).  
Based on the extent of the remnant as seen in the radio and X--ray 
observations it is likely that this emission is not part of \gsnr.
Substantial differences between the two mosaics are also seen in area G. 
Although the nature of the \oiii\ structure observed is filamentary all along 
the $\sim$42\arcmin\ of its length, the \oii\ emission appears quite faint and 
completely diffuse. 
Another area with great differences among the low and medium ionization 
images is area H. The structures are much better defined in \oiii\ than 
in \oii. The \oii\ image of the area seems to consist of several small scale 
filamentary structures and some of them do not have \oiii\ counterparts.
Finally, in area I we find that the \oiii\ emission is better structured 
compared to the \oii\ emission line image which, in addition, looks more  
clumpy. Note that around \a\ $\simeq$ 19\h41\m\ and \dd\ $\simeq$
32\degr20\arcmin\ faint \oiii\ emission is detected in an arc--like shape. 
Its typical brightness is $\sim$1/3 of the bulk of the emission seen a few 
arcminutes to the south.  
\section{Area C in detail}
Long observing runs would be required in order to map this 
remnant in the major optical emission lines. The only area observed 
in more detail is area C (Fig. \ref{fig03}) which was observed through the lines 
of \hnii, \sii, \oii\  and \oiii. 
These observations are flux calibrated and show that the emission, in all 
filters, is strong in absolute terms. However, there are striking differences 
between the low ionization lines and \oiii\ also in this area. 
The major difference being the double filament seen in the center of this 
field (Fig. \ref{fig03}). It is best defined in the medium ionization line 
of \oiii, while diffuse emission is present in the \oii\ image but is hardly 
detected in the images of \hnii\ and \sii. The two filaments are separated 
by $\sim$1\arcmin, while their typical projected thickness is less than 
$\sim$35\arcsec. This angular thickness corresponds to 0.17 pc, 
at a distance of 1 kpc, which is much smaller than the total length of this 
structure of $\sim$19 pc. We note here that only a small part 
($\sim$16\arcmin) of this filamentary structure is present in the low 
ionization images strongly suggesting the absence of complete recombination 
zones. This seems to be a common characteristic of the various areas of 
\gsnr. In order to explore the properties of the double filament, we formed the 
ratio of the \oii\ to the \oiii\ image. This new image could help to 
discriminate between projection effects and actual changes in the 
shock parameters. We find that the measured values of the ratio along each 
filament change by $\sim$20\%, while the average values of the ratios at 
the two filaments agree to within the 1.5$\sigma$ range. The relatively small 
difference in the ratio between the two filaments might point to simple 
projection effects but images of higher spatial resolution would be required 
to check this issue in more detail. 
\par
Finally, we also formed the ratio of the \oiii\ emission line image to the 
\hnii\ image which would allow to establish areas with complete and 
incomplete recombinations zones (Raymond \et\ \cite{ray88}). 
Indeed, this image clearly demonstrates that the 
double filament as well as the strong emission seen in the south--west are 
characterized by incomplete structures, while the emission from the south--east 
regions originates from complete shock structures.  

\section{The long--slit spectra from \gsnr}
We have obtained long--slit spectra at eight different locations of \gsnr. 
The absolute \ha\ flux covers a wide range of values from 0.5 to 22 $\times$
\flux\ (Table~\ref{sfluxes}). 
All spectra suggest that the emission detected during these spectroscopic 
observations originates from shock heated gas (\sii/\ha\ $\sim$0.7-1.8; 
Table \ref{sfluxes}). This conclusion also applies to the single spectrum taken 
from LBN 148 showing that this object is related to the remnant and is not 
an \HII\ region. The ratio of the sulfur lines emitted at 6716 \AA\ and 6731
\AA\ allow us to estimate the electron density (e.g. Osterbrock \cite{ost89}) 
at the position observed. The electron densities were calculated with the 
nebular package within the IRAF software (Shaw and Dufour \cite{sha95}). 
All ratios tend to the high end of the allowable range of values 
suggesting electron densities below $\sim$150 \dens\ and even 
below $\sim$30 \dens. Consequently, we cannot directly measure the 
preshock cloud density but can only place upper limits. 
The \ha/\hbeta\ ratio is used to estimate the logarithmic extinction c towards 
a source of line emission assuming an intrinsic ratio of 3. 
A signal to noise weighted average of the logarithmic extinction
c is 4.3 for the different locations around \gsnr. 
Higher values are found among the spectra but are also accompanied by 
larger errors. Fesen \et\ (\cite{fes85}) obtained spectra at two locations and
found \ha/\hbeta\  ratios around 3.3 ($\pm$ 35\%) in agreement with our 
measurements within the 1$\sigma$ range. 
\par
The most interesting result involves the very large \oiii/\hbeta\ ratios
observed in most of the spectra. Models of complete shock structures predict 
values below $\sim$6 for this ratio (Cox \& Raymond \cite{cox85}, 
Hartigan \et\ ~\cite{har87}). 
This limit is easily exceeded in case of shocks with incomplete 
recombination zones (Raymond \et\ \cite{ray88}). 
Our measured values for the  \oiii/\hbeta\  ratio range from 6 to 110 
(Table \ref{sfluxes}) indicating  that shocks with complete and incomplete 
recombination zones are present. Fesen \et\  (\cite{fes85}) measured this ratio to 
be $\sim$2 (their position 1) and $\sim$40 (their position 2).  
Our area 1 lies close to their position 2 ($\sim$90\arcsec apart) and there 
the measured ratios of \oiii/\hbeta\ are 46 and 86 along the slit (Table \ref{sfluxes}). 
These spectra were taken at different positions and thus, cannot be 
directly compared but they do point to the presence of incomplete recombination 
zones in this part of the remnant.
We note here that the signal to noise ratios quoted in Table \ref{sfluxes} 
do not include calibration errors which are less than $\sim$10\%. 
%

  \begin {figure*}
   \resizebox{\hsize}{!}{\includegraphics{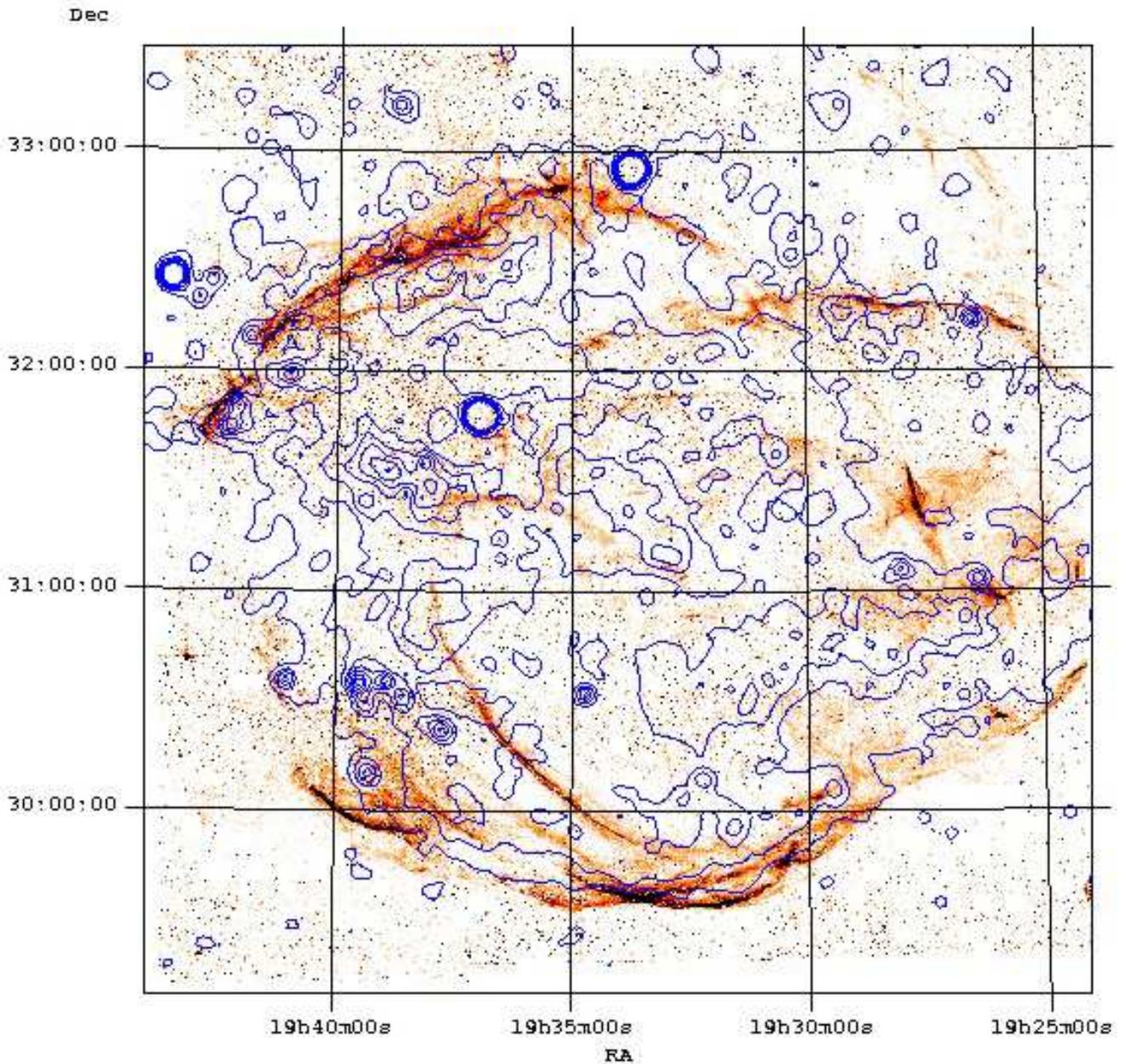}}
   \caption{ The \oiii\ emission of \gsnr\ shown together with the soft X--ray 
   emission detected from ROSAT (Lu \& Aschenbach \cite{lu02}). 
   The X--ray emission is shown in contours 
   starting from 2.5 10$^{-4}$ cts s$^{-1}$ pixel$^{-1}$ in steps of 
   3.5 10$^{-4}$ cts s$^{-1}$  pixel$^{-1}$, while each pixel is 
   0\arcmin.75 $\times$ 0\arcmin.75.  } 
     \label{fig04}
   \end{figure*}
%
\par
As is evident from Table \ref{sfluxes} only a small number of the extracted
spectra points to shocks with complete recombination zones. 
The spectra from positions 1, 2, 3a, 4, 6, 7, and 8 strongly support the 
presence of incomplete recombination zones (\oiii/\hbeta\ $\sim$10--100). 
Examination of the theoretical estimates of the \oiii/\hbeta\ and \oi /\hbeta\ 
ratios as a function of column density (Raymond \et\ \cite{ray88})  
shows that shock speeds higher than $\sim$100 \vel, most likely 
around 120 \vel, and column densities in the range of 10$^{17}$ -- 10$^{18.5}$ 
\sdens\ are required to match the observations.  
Especially, at position 1 a shock velocity of 120 \vel\ and a column density of
$\sim$10$^{17.5}$ \sdens\ is estimated, which is well below that of  
$\sim$10$^{18.7}$ \sdens\  required for a complete 
recombination zone. 
\section{Discussion}
The extended remnant \gsnr\ was observed for the first time with a highly 
efficient CCD camera in the low and medium ionization lines of \oii 3727 \AA\ 
and \oiii 5007 \AA, respectively. The \oii\ interference filter being in the far
blue part of the optical wavelength band provides sharp images since the
emission from field stars is significantly reduced, while any morphological
differences between the two images can not be attributed to abundance
variations. 
In addition, we performed the first flux calibrated imaging observations of 
a specific area in the south of \gsnr\ (area C, \S 4) in major optical 
emission lines. These observations as well as the raster observations reveal 
in sufficient detail what is already known for parts of the remnant, i.e. strong
differences between the lower and higher ionization lines 
(Rosado \cite{ros81}, Fesen \et\ \cite{fes83}, Sitnik \et\ \cite{sit83}).
\par
The object under study is an excellent example of the class of remnants 
displaying prominent \oiii\ emission and is this line that provides 
the sharpest view of the system (e.g. Fesen \et\ \cite{fes97}, 
Mavromatakis \et\ \cite{mav02}). 
In the following we will assume a distance 
of 1 kpc to the remnant and an angular radius of 1\degr.8 implied by the 
X--ray (Lu \& Aschenbach \cite{lu02}) and our optical data. 
The thinnest filamentary structures seen in the \oii\ images are characterized 
by typical projected widths, actually the full width at half maximum (fwhm), 
of $\sim$20\arcsec\ or 0.1 pc, while the
shortest filaments are $\sim$0.5 pc long. The longest ones are of the 
order of several pc. We also find gaps in certain filaments which otherwise 
look continuous. The shortest gaps are typically around 0.2 pc suggesting 
this length as the typical length of small scale inhomogeneities in this 
cloudy environment.  
\par
A shock velocity of $\sim$120 \vel\ and a column density of $\sim$10$^{17.5}$
--10$^{18}$ \sdens\ are estimated for area 1 from the optical spectra and the 
modeling results of Raymond \et\  (\cite{ray88}).   
The measured fwhm of 0.14 pc of the filament in this area and the estimated 
column density allow us to estimate an average filament density of 
$\sim$0.8--2.5 \dens. The actual preshock cloud density must be lower 
than this by, at least, a factor of 4. 
An IUE spectrum taken a few arcseconds off our position
shows emission from \ion{C}{iii}], \ion{N}{iii}], \ion{O}{iii}], but weak \ion{C}{iv} 
and maybe \ion{Si}{iv}, and \ion{O}{iv} 
(Raymond J. 2002, private communication). Thus, it is not possible to
fully explore the physical conditions in this area as Raymond \et\
(\cite{ray88}) have done for a bright filament in the \object{Cygnus Loop}.  
Using the same approach to other areas of the remnant we find average densities 
in the range of $\sim$0.3--4 \dens. 
Specifically, for area 5 a shock velocity of 90 \vel\ and column densities in 
the range of 10$^{18}$ -- 10$^{18.5}$ \sdens\  are estimated.
Extraction of the intensity profiles from the \oii\ and \oiii\ images along the 
declination axis shows that the positions of maximum intensity differ by 
$\sim$0.07 pc with the \oiii\ profile leading, while the fwhm of the fitted 
Gaussian profiles are 0.13 and 0.19 pc, respectively. 
This behaviour of the lower and higher ionization lines is also verified by the 
long slit spectra which are characterized by a spatial scale of 1/350 pc 
compared to that of 1/40 pc of the wide field images. 
Assuming a typical length of 0.19 pc and a column density of $\sim$10$^{18.3}$ 
\sdens, we estimate the average density behind the radiative shock as 
$\sim$3.5 \dens. These simplified calculations show that in several areas of 
the remnant the particle density in the recombination region is of the order
of a few atoms per cm$^3$.
\par
In areas where complete recombinations zones are suggested by the observational 
data we can utilize the relations  
\begin{equation}
{\rm n_{[\ion{S}{ii}]} \simeq\ 45\ n_c \ V_{s}^2},
\end{equation}
and
\begin{equation}
{\rm E_{51} = 2 \times 10^{-5} \beta^{-1}\ n_c\  V_{s}^2\ r_{s}^3},  
\end{equation}

given by Fesen \& Kirshner (\cite{fes80}) and McKee \& Cowie 
(\cite{mck75}), respectively, to limit the preshock cloud density and explosion 
energy. The factor $\beta$ is of the order of 1--2, 
${\rm n_{[\ion{S}{ii}]}}$ is the electron density derived from the sulfur 
line ratio, n$_{\rm c}$ is the preshock cloud density, ${\rm E_{51}}$ is the
explosion energy in units of 10$^{51}$ erg, V$_{\rm s}$ is the shock velocity
into the clouds in units of 100 \vel, and  r$_{\rm s}$ is the radius of the 
remnant in pc. 
Assuming an upper limit of $\sim$70 \dens\ on the electron density 
(Table 2; areas 3b, 4, 5) we find that ${\rm n_c V_s^2}$ $<$ 1.6 and 
consequently, ${\rm E_{51}}$ $<$ 1 for r$_{\rm s}$ $\simeq$ 32 and 
$\beta \simeq$ 1. Although this limit is rather broad, it is consistent with the 
explosion energy of 0.2 estimated by Lu \& Aschenbach (\cite{lu02}) from the 
Sedov--Taylor solution of the X--ray data. 
However, in order to derive Eqs. (1) and (2) it is assumed that the magnetic
field component perpendicular to the shock velocity vector is negligible 
(see also Bocchino \et\ \cite{boc00}). If
this is not the case then the preshock cloud density will be expected to be 
higher than what Eq. (1) implies. Comparison of the shock velocities
estimated from the optical data with the shock velocity of the primary blast 
wave suggests density contrasts $\sim$10--20 between the interstellar medium and the 
interstellar clouds. The preshock medium density of 0.02 \dens\ given by 
Lu \& Aschenbach (\cite{lu02}) implies typical preshock cloud densities 
$\sim$0.2--0.4 \dens, which are consistent with the upper limits obtained earlier. 
\par
The observed (\ha/\hbeta)$_{\rm obs}$ ratio can be used to estimate the 
logarithmic extinction c towards a source of line emission. 
The relation we use is c = 1./0.331 * log((\ha/\hbeta)$_{\rm obs}$/3.) and 
is based on the interstellar extinction curve of Kaler (\cite{kal76}) as 
implemented in the nebular package (Shaw and Dufour (\cite{sha95}) within 
the IRAF software.  The minimum extinction of 0.24 ($\pm$0.04)
is measured in area 5, while the maximum, statistically significant 
measurement of 0.8 ($\pm$0.2) is found in area 1a. 
The absorption at area 1 and
2 is higher than in other parts of the remnant. This may be related to the 
presence of excess emission seen in the IRAS 60 $\mu$m maps 
(Wheelock \et\ \cite{whe94}) to the east 
of 19\h40\m\ provided that the dust emission lies to the foreground of \gsnr. 
The observed variations of the logarithmic extinction may indicate the 
presence of intrinsic absorption although these variations are at the 3--4 
$\sigma$ level. Note that variations in the column density, 
determined from X--ray measurements, are reported by Lu \& Aschenbach 
(\cite{lu02}). The probable non--uniform distribution of the column 
density and the logarithmic extinction would suggest a complex, 
inhomogeneous environment into which the primary blast wave propagates.
In order to estimate the column density implied by the optical spectra 
we use the relation N$_{\rm H}$ = 5.4($\pm$0.1) $\times$~$10^{21}$ E(B-V)
\sdens\ mag$^{-1}$ (Predehl and Schmitt \et\ \cite{pre95}). 
Assuming a logarithmic extinction of 0.36($\pm$0.03), the average from 
areas 3, 4, and 5, which is equivalent to an E(B-V) of 0.24 
(E(B--V) = 0.664 c ; Kaler \cite{kal76}, Aller \cite{all84}), 
a column density of 13($\pm1$) $\times$10$^{20}$ \sdens\ is calculated. 
This column density exceeds that of 3--5$\times$10$^{20}$ \sdens\  
measured 
in the ROSAT pointed data (Lu \& Aschenbach \cite{lu02}) by factors in the 
range of 2.6 - 4.3. It is not clear at the moment if this implies the 
presence of excess intrinsic absorption affecting mainly the optical 
wavelengths or is due to a statistical effect since these factors are 
just compatible with the scattering around the straight line fit of 
Predehl and Schmitt (\cite{pre95}).
\par
The supernova remnant \gsnr\ was observed in the soft X--ray band by ROSAT
through a large number of pointed observations as well as during the
All--Sky survey (Lu \& Aschenbach \cite{lu02}). 
In Fig. \ref{fig04} we show the \oiii\ emission together with the contours 
of the soft X--ray emission detected by ROSAT. 
The outer \oiii\ filaments are rather well correlated with the X--ray emission 
mainly in the south, and north, north--east areas. The southern \oiii\ 
filaments probably lie close to the primary shock wave, while X--ray emission 
is detected ahead of the northern \oiii\ filaments. The clumpiness of the 
X--ray emission in the central areas of the remnant does not allow a reliable 
comparison with the optical emission. The temperature or column density maps do
not provide any clear evidence for a correlation or anticorrelation with the 
optical data. There is also no well defined X--ray emission outside the shell 
of the remnant around \a\ $\simeq$ 19\h28\m, \dd\ $\simeq$ 33\degr00\arcmin\
suggesting that the detected optical emission may not be related to it.

\section{Conclusions}
The supernova remnant \gsnr\ was observed in the lower and 
medium ionization lines of \oii 3727 \AA\ and \oiii 5007 \AA. 
The \oiii\ emission line mosaic is more filamentary than the \oii\ image. 
Several new structures are detected, while 
some of them are found outside the X--ray shell indicating that those may 
not be related to \gsnr. The morphology of the \oii\ and \oiii\ images suggest the 
presence of small scale inhomogeneities which are of the 
order of $\sim$0.2 pc. Larger scale inhomogeneities are also present. 
The long slit spectra clearly demonstrate the very strong 
\oiii\ emission relative to \hbeta\  suggesting the presence of incomplete
recombination zones. The shock velocities are found in the range of 90--140
\vel, while the column densities behind the radiative shock cover the range 
of $\sim$ 10$^{17.0} - 10^{18.5}$ \sdens. The extinction variations seen 
in the optical spectra may indicate the presence of internal absorption.
\begin{acknowledgements}
The authors would like to thank J. Raymond and the referee B. Aschenbach  
for their useful comments and suggestions.
Skinakas Observatory is a collaborative project of the University of
Crete, the Foundation for Research and Technology-Hellas and
the Max-Planck-Institut f\"ur Extraterrestrische Physik. This research has 
made use of data obtained through the High Energy Astrophysics Science 
Archive Research Center Online Service, provided by the NASA/Goddard 
Space Flight Center.
\end{acknowledgements}
%



\begin{thebibliography}{}
	\bibitem[1984]{all84} Aller L. H. 1984, ``Physics of thermal gaseous
		nebulae'', D. Reidel Publishing Company

	\bibitem[1994]{asc94} Aschenbach B. 1994,  ROSAT observations 
		of supernova remnants, in Proc. of ``New Horizon of X--ray
		Astronomy'', eds. F. Makino \& T. Ohashi (Universal Academy
		Press), p. 103
	
	\bibitem[2000]{boc00} Bocchino F., Maggio A., Sciortino S. and Raymond
		J. 2000, A\&A 359, 316
	
	\bibitem[1985]{cox85} Cox D. P., Raymond J. C. 1985,
		ApJ 298, 651

        \bibitem[1980]{fes80} Fesen R. A., Kirshner R. P. 1980, 
     		ApJ 242, 1023

        \bibitem[1983]{fes83} Fesen R. A., Gull T. R. and Ketelsen D. A. 1983, 
		ApJS 51, 337

        \bibitem[1985]{fes85} Fesen R. A., Blair W. P. and Kirshner R. P. 
                1985, ApJ 292, 29

        \bibitem[1997]{fes97} Fesen R. A., Winkler P.F., Rathore Y., 
                Downes R.A., Wallace D., Tweedy R.W. 1997, 
                AJ 113, 767  

        \bibitem[1996]{gor96} Gorham P. W., Ray P. S., Anderson S. B., 
                Kulkarni S. R. and Prince T. A. 1996, ApJ 458, 257

        \bibitem[1977]{gul77} Gull T.R., Kirshner R. P. and Parker R. A. R.
                1977, ApJ 215, L69

	\bibitem[1987]{har87} Hartigan P., Raymond J. and Hartmann L. 1987, 
		ApJ 316, 323
	
	\bibitem[1976]{kal76} Kaler J. B. 1976, ApJS 31, 517
	
	\bibitem[1999]{las99} Lasker B. M., Russel J. N. \& Jenkner H., 1999,
                in the HST Guide Star Catalog, version 1.1-ACT, The
                Association of Universities for Research in Astronomy, Inc

	\bibitem[2002]{lu02} Lu F. J. and Aschenbach B. 2002, to be submitted
	
        \bibitem[1965]{lyn65} Lynds B. T. 1965, ApJS 12, 163  

	\bibitem[1979]{mas79} Mason K. O., Kahn S. M., Charles P. A., Lampton 
		M. L. and Blisset R. 1979, ApJ 230, L163
		 
	\bibitem[2000]{mav00} Mavromatakis F., Papamastorakis J., Paleologou 
		E. V., and Ventura J. 2000, A\&A 353, 371

	\bibitem[2002]{mav02} Mavromatakis F., Boumis P. Paleologou E. V. 2002, 
		A\&A 383, 1011
		
	\bibitem[1975]{mck75} McKee C. F., Cowie L. 1975,
		ApJ 195, 715
	
        \bibitem[1989]{ost89} Osterbrock D. E. 1989,  Astrophysics of 
	           gaseous nebulae, W. H. Freeman \& Company  
     
	\bibitem[1995]{pre95} Predehl P., and Schmitt J. H. M. M. 1995, 
		A\&A 293, 889
		
	\bibitem[1988]{ray88} Raymond J. C., Hester J. J.,
                Cox. D., Blair W. P., Fesen R. A., Gull T. R. 1988, 
                ApJ 324, 869
		
        \bibitem[1979]{rei79} Reich W., Kallas E. and Steube R. 1979,
                A\&A 78, L13

        \bibitem[1981]{ros81} Rosado M. 1981, ApJ 250, 222
    	
        \bibitem[1976]{sab76} Sabbadin F. and D'Odorico S. 1976, A\&A,
                49, 119

        \bibitem[1990]{sew90} Seward F. 1990, ApJS, 73, 781

	\bibitem[1995]{sha95} Shaw R. A. and Dufour R. J. 1995, PASP 107, 896
	
	\bibitem[1983]{sit83} Sitnik T. G., Klement\'eva A. Y. and Toropova M.
                 S. 1983, Sov.Astr. 27, 292
	
	\bibitem[1978]{sny78} Snyder W. A., Davidsen A. F., Henry R. C.,
		Shulman S., Fritz G., and Friedman H. 1978, ApJ 222, L13

	\bibitem[1994]{whe94}  Wheelock S. L., Gautier T. N., Chillemi J.,
                Kester D., Mccallon H., Oken C., White J., Gregorich D.,
                Boulanger F., Good J. 1994,
                IRAS sky survey atlas, Explanatory supplement
%
\end{thebibliography}
\end{document}